\documentclass[draft]{agujournal2019}
\usepackage{url} %this package should fix any errors with URLs in refs.
\usepackage{lineno}
\usepackage[inline]{trackchanges} %for better track changes. finalnew option will compile document with changes incorporated.
\usepackage{soul}
\usepackage{amsmath}
\usepackage{algorithm}
\usepackage{algpseudocode}
\usepackage{booktabs}
\usepackage{natbib}
\draftfalse

\journalname{Journal of Advances in Modeling Earth Systems (JAMES)}

\begin{document}

\pdfminorversion=5
\pdfcompresslevel=9
\pdfobjcompresslevel=3
\pdfimageresolution=150

%% ------------------------------------------------------------------------ %%
%  Title
%
% (A title should be specific, informative, and brief. Use
% abbreviations only if they are defined in the abstract. Titles that
% start with general keywords then specific terms are optimized in
% searches)
%
%% ------------------------------------------------------------------------ %%

% Example: \title{This is a test title}

\title{DiffESM: Conditional Emulation of Temperature and Precipitation in Earth System Models\\with 3D Diffusion Models}

%% ------------------------------------------------------------------------ %%
%
%  AUTHORS AND AFFILIATIONS
%
%% ------------------------------------------------------------------------ %%

% Authors are individuals who have significantly contributed to the
% research and preparation of the article. Group authors are allowed, if
% each author in the group is separately identified in an appendix.)

% List authors by first name or initial followed by last name and
% separated by commas. Use \affil{} to number affiliations, and
% \thanks{} for author notes.
% Additional author notes should be indicated with \thanks{} (for
% example, for current addresses).

% Example: \authors{A. B. Author\affil{1}\thanks{Current address, Antartica}, B. C. Author\affil{2,3}, and D. E.
% Author\affil{3,4}\thanks{Also funded by Monsanto.}}

\authors{Seth Bassetti\affil{1}\thanks{Work performed while at Western Washington University.}, Brian Hutchinson\affil{2,3}, Claudia Tebaldi\affil{4}, Ben Kravitz\affil{5,6}}

\affiliation{1}{Computer Science Department, Utah State University, Logan, UT}
\affiliation{2}{Computer Science Department, Western Washington University, Bellingham, WA}
\affiliation{3}{Foundational Data Science, Pacific Northwest National Laboratory, Seattle, WA}
\affiliation{4}{Joint Global Change Research Institute, Pacific Northwest National Laboratory, College Park, MD}
\affiliation{5}{Earth and Atmospheric Sciences Department, Indiana University, Bloomington, IN}
\affiliation{6}{Atmospheric Sciences and Global Change Division, Pacific Northwest National Laboratory, Richland, WA}
%(repeat as many times as is necessary)

%% Corresponding Author:
% Corresponding author mailing address and e-mail address:

% (include name and email addresses of the corresponding author.  More
% than one corresponding author is allowed in this LaTeX file and for
% publication; but only one corresponding author is allowed in our
% editorial system.)

% Example: \correspondingauthor{First and Last Name}{email@address.edu}

\correspondingauthor{Brian Hutchinson}{Brian.Hutchinson@wwu.edu}

%% Keypoints, final entry on title page.

%  List up to three key points (at least one is required)
%  Key Points summarize the main points and conclusions of the article
%  Each must be 140 characters or fewer with no special characters or punctuation and must be complete sentences

% Example:
% \begin{keypoints}
% \item	List up to three key points (at least one is required)
% \item	Key Points summarize the main points and conclusions of the article
% \item	Each must be 140 characters or fewer with no special characters or punctuation and must be complete sentences
% \end{keypoints}

\begin{keypoints}
\item Earth system models (ESMs) are key devices for understanding how human actions will affect the future global climate.
%\item Generating a sufficient number of samples with an ESM is computationally demanding, and often intractable for analyses requiring large statistical power, like the characterization of extreme events.
\item Computational demands prevent us from running them for more than a handful of scenarios. Consequently, ESM emulators are often limited to monthly frequency.

\item We present DiffESM as a data-driven emulator of ESMs that closely matches the spatiotemporal distributions of ESMs at daily frequency.
\end{keypoints}

%% ------------------------------------------------------------------------ %%
%
%  ABSTRACT and PLAIN LANGUAGE SUMMARY
%
% A good Abstract will begin with a short description of the problem
% being addressed, briefly describe the new data or analyses, then
% briefly states the main conclusion(s) and how they are supported and
% uncertainties.

% The Plain Language Summary should be written for a broad audience,
% including journalists and the science-interested public, that will not have 
% a background in your field.
%
% A Plain Language Summary is required in GRL, JGR: Planets, JGR: Biogeosciences,
% JGR: Oceans, G-Cubed, Reviews of Geophysics, and JAMES.
% see http://sharingscience.agu.org/creating-plain-language-summary/)
%
%% ------------------------------------------------------------------------ %%

%% \begin{abstract} starts the second page

\begin{abstract}
Earth System Models (ESMs) are essential for understanding the interaction between human activities and the Earth's climate. However, the computational demands of ESMs often limit the number of simulations that can be run, hindering the robust analysis of risks associated with extreme weather events. While low-cost climate emulators have emerged as an alternative to emulate ESMs and enable rapid analysis of future climate, many of these emulators only provide output on at most a monthly frequency. This temporal resolution is insufficient for analyzing events that require daily characterization, such as heat waves or heavy precipitation. We propose using diffusion models, a class of generative deep learning models, to effectively downscale ESM output from a monthly to a daily frequency. Trained on a handful of ESM realizations, reflecting a wide range of radiative forcings, our DiffESM model takes monthly mean precipitation or temperature as input, and is capable of producing daily values with statistical characteristics close to ESM output. Combined with a low-cost emulator providing monthly means, this approach requires only a small fraction of the computational resources needed to run a large ensemble. We evaluate model behavior using a number of extreme metrics, showing that DiffESM closely matches the spatio-temporal behavior of the ESM output it emulates in terms of the frequency and spatial characteristics of phenomena such as heat waves, dry spells, or rainfall intensity. 
\end{abstract}

\section*{Plain Language Summary}
Ideally, to study how damaging phenomena like heatwaves, droughts and downpours will change in the future under global warming, we would want a large number of climate model runs producing many realizations of climate futures that we can analyze and from which the new characteristics of climate extremes can be quantified. Currently, emulators can rapidly generate simulations of future climate, but often to relatively low frequencies, as decadal, annual or monthly output at best in most cases, which is insufficient for studying extreme events that occur on a daily timescale. We show how it is possible to train a machine learning model to produce daily series of temperature or precipitation from monthly averages, thus facilitating  a more robust investigation into how extreme events will change in the future.

%% ------------------------------------------------------------------------ %%
%
%  TEXT
%
%% ------------------------------------------------------------------------ %%

%%% Suggested section heads:
% \section{Introduction}
%
% The main text should start with an introduction. Except for short
% manuscripts (such as comments and replies), the text should be divided
% into sections, each with its own heading.

% Headings should be sentence fragments and do not begin with a
% lowercase letter or number. Examples of good headings are:

% \section{Materials and Methods}
% Here is text on Materials and Methods.
%
% \subsection{A descriptive heading about methods}
% More about Methods.
%
% \section{Data} (Or section title might be a descriptive heading about data)
%
% \section{Results} (Or section title might be a descriptive heading about the
% results)
%
% \section{Conclusions}

\section{Introduction}

Extreme weather events, such as heat waves, droughts, and floods have become more frequent and intense in recent years \citep{Seneviratneetal}. These events have significant impacts on human societies and ecosystems, highlighting the urgent need to understand how they may change in the future under different emission scenarios. One important tool for investigating future climate change and its impacts on extreme weather events is the use of Earth System Models (ESMs) run under plausible future scenarios of greenhouse gas emissions.

ESMs are complex computer models that simulate the interactions between Earth's atmosphere, oceans, land surface, biosphere, and cryosphere. They are used to simulate a wide range of climate variables under different emissions scenarios. However, the computational demands of ESMs limit the number of simulations that can be performed, especially when climate modeling centers need to allocate experiments to meet demands from a range of scientific and practical uses. This is especially problematic when investigating rare extreme weather events, as it is necessary to aggregate data over numerous runs to obtain reliable statistics. To address this issue, emulators can be used to generate realizations of global climate data in the scale of minutes or hours, rather than weeks or months \citep{kasim2021building, stitches, nath2021mesmer}. Emulators learn the statistical characteristics of ESM output from existing data, and can then generate new data under the same scenario used for training but also, importantly, under different emissions scenarios, only utilizing some type of simplified, coarser-scale ``conditioning'' from the target scenario. Current emulators, such as MESMER-M \citep{nath2021mesmer} have been shown to have good performance in generating fields of variables at monthly frequency. However, to fully investigate certain extreme events, such as heat waves, dry spells, or intense precipitation, a daily frequency is necessary. This can be efficiently addressed by a temporal downscaling mechanism to produce daily values from monthly averages. Machine learning approaches, especially generative deep learning methods have shown promising results as spatial downscaling models,  \citep{latent-diffusion, imagen, base-denoising}, and we extend these techniques to temporal downscaling mechanisms.

In this paper, we present a denoising diffusion probabilistic model which learns to closely model the spatio-temporal behavior of an ESM, producing month-long samples of either daily mean temperature or precipitation. Diffusion models have shown great success in the realm of generative modeling, extending to the domains of image, audio, video, and recently even climate. Our emulator, DiffESM, can be used to complement current emulators such as fldgen \citep{fdlgen}, MESMER \citep{Beuschetal,Nathetal, Quilcailleetal}, by producing series of daily quantities consistent with these emulators' monthly means. 

Thus, once trained, by combining DiffESM with existing  emulators that provide monthly output, researchers can rapidly investigate the effect of climate scenarios on the distribution and characteristics of extreme weather events at a daily timescale similarly to what could be done if an ESM were to be run under those scenarios in a ``large ensemble'' mode~\citep{Deseretal2020}.  By enabling the rapid analysis of extreme event statistics for a range of scenarios and ESMs, DiffESM can provide valuable insights into the range and magnitude of potential climate impacts, and help inform adaptation and mitigation strategies.

\section{Related Work}
In recent years, machine learning has gained increased attention as a tool to support research in the earth sciences \citep{deep-learning-earth-science, rolnick2019tackling}. One promising area of application for data-driven algorithms is in forecasting. Traditional weather forecasting relies on physically-constrained models, which can be computationally expensive. Neural networks have emerged as a promising alternative to traditional forecasting models, demonstrating the ability to emulate complex physical processes at a significantly lower computational cost \citep{forecasting1, forecasting2, forecasting3, forecasting-4}. Consequently, machine learning algorithms are increasingly being considered as a complement or even standalone approach to weather forecasting.

One such example is the use of generative models for forecasting. Traditionally used for image generation, these models are able to approximate complex data distributions and generate new samples from those distributions. Generative adversarial networks (GANs), a class of generative models that rely on a generator-discriminator architecture \citep{gans}, have emerged as useful forecasting tool due to their ability to generate spatially coherent data. \citep{gan-fore-1, gan-fore-2}. More recently, diffusion models have surpassed GANs in terms of performance \citep{diff-beat-gans} and are currently being explored for applications including weather forecasting \citep{chen2023swinrdm, price2023gencast}, solar irradiance forecasting \citep{solar-forecasts}, precipitation now-casting \citep{latent-nowcast, gao2024prediff}, emulating forecast ensembles \citep{seeds} and beyond.

Outside of weather forecasting, data-driven approaches have found applications in conjunction with ESMs and Regional Climate Models (RCMs). These models often have limited spatial grid sizes, typically 100km or more for the former, while the latter can be run for limited domains at higher resolution but are still very burdensome, computationally. To facilitate the provision of climate information at fine scales, machine learning models have been employed to spatially downscale the output of ESMs and RCMs, enabling the generation of higher-resolution projections \cite{downscaling-1, downscaling-2, downscaling-3}. Furthermore, the coarse spatial grid sizes of climate models pose challenges in simulating sub-grid scale processes, such as cloud formation. In this context, machine learning models have been utilized to emulate these processes, as an alternative to more traditional parameterization choices for their incorporation into climate simulations \citep{cloud-1, cloud-2}.

The computational complexity of climate models poses a significant challenge in climate science. One recent study on the performance of climate models states that the IPSL-CM5A-LR,  which is the focus of this work, takes a full real-time day to generate just 6 years of data \citep{cpmip}. These computational limitations hinder the use of climate model outputs in impact research, where ideally, climate information under arbitrary scenarios would be readily available. To address this challenge, various approaches to climate model emulation have been explored. One such approach is the use of Reduced Complexity Models, computationally efficient tools which typically have a much lower resolution than ESMs (often as low as global mean scales)  \citep{reduced-complexity-1, reduced-complexity-2}. In addition, generative modeling techniques have shown promising results in emulating the spatio-temporal behavior of climate models, outperforming more conventional statistical approaches such as \citep{stat-emulation-1, stat-emulation-2}.  Generative Adversarial Networks (GANs) have been extensively used for this purpose \citep{clim-gan-1, clim-gan-2, clim-gan-3, clim-gan-4}, but recently, there is growing interest in exploring the use of diffusion models for this task \citep{diff-esm, uk-diff}.

\section{Background}
\subsection{Discrete Time Diffusion}
Deep generative models are a type of generative model that use deep learning techniques such as neural networks to generate data that is similar to real-world data. Unlike traditional generative models, deep generative models are able to generate data with a much higher level of detail, accuracy, and complexity. In essence, deep generative models are capable of learning the underlying patterns and distributions of the data, allowing them to generate highly realistic data samples. 

Denoising Diffusion Probabilistic Models (DDPM), or simply diffusion models, are a class of generative models that are both flexible and tractable, which learns to transform a sample from a known distribution, such as a Gaussian, into a sample that could be drawn from an unknown distribution \citep{og-diffusion, base-denoising, improved-diffusion}. 

Diffusion models are trained by systematically destroying information in a sample and learning to reconstruct it. This sample is destroyed with an iterative forward diffusion process over a series of time-steps, starting from $x_0$ (samples from the data distribution) to $x_T$ (Guassian noise). The inspiration for these models was drawn from non-equilibrium statistical physics. The noised sample at a given time-step ($x_t$) depends on the noised sample at the previous time-step, with conditional probability defined as follows:

\begin{equation}
q(x_t|x_{t-1})=\mathcal{N}(x_t; \sqrt{1-\beta_t}x_{t-1},\beta_tI)
\end{equation}
where $\beta$ is a fixed noise schedule chosen beforehand. One common example for the noise schedule is a linearly spaced range from $\beta_0 = 0.0001$ to $\beta_{999} = 0.02$.

However, because the addition of normal distributions results in a normal distribution, the amount of noise at a given noise-scale from a sample with zero noise can be calculated without needing to calculate each noisy sample in between. The updated conditional probability is defined below:
\begin{equation}
    \alpha_t=1-\beta_t
\end{equation}
\begin{equation}
    \bar{\alpha}_t=\prod_{i=0}^{T-1}\alpha_i
\end{equation}
\begin{equation}
    q(x_t|x_0)=\mathcal{N}(x_t;\sqrt{\bar{\alpha}}x_0,\sqrt{1-\bar{\alpha}})
\end{equation}

An alternative way to define the conditional probability above is:
\begin{equation}
    \epsilon\sim\mathcal{N}(0, \mathcal{I})
\end{equation}
\begin{equation}
        q(x_t|x_0)=\sqrt{\bar{\alpha}}x_0 + \sqrt{1-\bar{\alpha}}  \, \epsilon
\end{equation}

The reverse process is a Markov chain which converts the sample from a known distribution (Gaussian) into a sample from the unknown distribution. Inference involves making small denoising steps, based on estimates of the cumulative noise; i.e., sampling $x_{t-1}$ from $x_t$ given an estimate of $x_0$ from $x_t$. Generating as a sequence of iterative steps is easier than having to model the sample distribution directly (i.e., mapping directly from $x_T$ to $x_0$). The loss for each denoising step is simply the mean-squared error between the original noise added to a sample ($\epsilon$) and the predicted noise ($\epsilon_\theta$) added to the sample. For further information on discrete time diffusion, we refer readers to \citep{base-denoising, improved-diffusion, diff-beat-gans}. 

In our work, rather than predict the original noise directly, we elect to predict a reparameterization of $\epsilon$, referred to as $v$. Empirically, this has improved our results and the technique was introduced in \citep{salimans2022progressive}.

\section{Methods}
\subsection{Setup}
Our objective is to develop a diffusion model capable of effectively downscaling monthly input data into target daily values. It is essential to note that this model is non-deterministic, which means that we are not aiming to replicate the exact sequence of days from our conditioning data. Instead, our goal is to create a random sequence of daily data (or an ensemble of them) that could have resulted in the monthly average used for conditioning, and preserves the spatio-temporal characteristics of daily sequences generated by the ESM.
By taking this approach, we aim to generate a distribution of physically-plausible weather samples. These generated samples should align with key metrics from the true conditioning sequence, such as the occurrence of hot streaks or intense precipitation events. This alignment ensures that the downscaled daily values maintain a realistic representation of the weather patterns observed in the monthly data. Below we describe the data we train our model on, the model architecture, and our scheme for training and inference.
\subsection{Data}
Our work focuses on two CMIP5-era \citep{Tayloretal} datasets made up of initial condition ensemble members, which we refer to as ``realizations,'' from the Community Earth System Model (CESM1-CAM5) \citep{cesm} and the Institut Pierre-Simon Laplace Earth System Model (IPSL-CM5A-LR) \citep{ipsl}. Separate diffusion models, emulating CESM and IPSL, respectively, are trained using four ensemble members, while two additional ensemble members are reserved for evaluation purposes (these will be labelled Held Out 1 and Held Out 2 in the following).  We bilinearly interpolate the CESM dataset to a spatial grid of 96 x 96 pixels, the same spatial resolution as the IPSL dataset. All realizations consist of daily average temperature (originally Kelvin transformed into degree Celsius) or total precipitation (mm/day) output from 1850 - 2100 for IPSL and 1920 - 2100 for CESM. Since we are interested in generating a ``month'' of data with our model, we train the diffusion model using datasets that have been segmented into overlapping 28-day chunks. This has the advantage of representing regular four-week periods. Due to the shorter year range covered by the CESM dataset (1920-2100), it consists of 264,152 training samples, whereas the IPSL dataset contains 366,352 training samples.
We train our model using both historical data as well as output from the highest-emission scenario available (RCP8.5 \citep{Mossetal}) in order to expose the emulator to a wide range of forcing levels. We then evaluate our model on unseen scenarios (specifically RCP2.6 and RCP4.5) to test its ability to generalize to different scenarios of anthropogenic forcing. 

\subsection{Data Pre- and Post-processing}
Like many machine learning methods, diffusion models operate best when working with data normalized to a certain range.  To achieve this, we apply several preprocessing and normalization steps to the temperature and precipitation data. For temperature, we convert the units from Kelvin to Celsius and then standardize the results using the overall mean and standard deviation of the training set. This standardization helps to center the data around zero and scale it to a consistent range, which we find is beneficial for model training. 

For precipitation, we apply a cubed root normalization to each value. This helps to shrink the long tail of extreme precipitation values and ensure that days with zero or very small amounts of precipitation (which happen frequently in many areas of the world in a climate model) fall within a regime that is more suitable for training. 

As a post-processing step for precipitation, prior to computing metrics, we set values below 1mm/day (i.e., dry days) to exactly 0mm/day, a common threshold choice to circumvent the tendency of ESMs to ``drizzle'' too frequently.

\subsection{Model Architecture}
Diffusion based-approaches have seen much recent success in the realm of video generation. In this work, we construct our own model architecture that is highly inspired by the Video Diffusion \citep{video-diffusion} architectures. Specifically, our denoising model is a 3D U-Net similar to the architecture described in \citealt{og-unet}. Each downsampling and upsampling stage in our U-Net is composed of multiple spatial-only convolutions, followed by a single temporal-only convolutional block. This allows our model to separate learning spatial structures and temporal structures, both of which are vital to creating realistic climate sequences. 

\subsection{Model Training and Inference}
Our training algorithm is described in Algorithm \ref{alg:model_training}.
For our diffusion processes, we use a linearly spaced noise schedule with 1000 time-steps. For inference, we use the DPMSolver++ algorithm \citep{lu2022dpm}, and 25 evenly spaced timesteps between 0 and 1000 for sampling. Additionally, we implement an exponential moving average (EMA) strategy for weight updates during training  \citep{song2020improved}. We use an exponential moving average factor of 0.999 to update our weights, which stabilizes the model weights for inference. We train each model for 5 epochs, stopping early upon a plateau in our loss curves. Using four Nvidia Titan V GPUs, total training time for a model typically spans two-three days.
\begin{algorithm}[t]
\caption{Training Denoising Diffusion Probabilistic Models (DDPM)}
\label{alg:model_training}
\begin{algorithmic}[1]
    \State Initialize model parameters $\theta$
    \State Sample random month from the dataset, $x_0$, of shape [time $\times$ latitude $\times$ longitude]
    \State Take the average of $x_0$ over the time dimension to create a conditioning map, $c$ of shape [latitude $\times$ longitude]
    \State Sample random noise $\epsilon$ ([time $\times$ latitude $\times$ longitude]) from $\mathcal{N}(0, \mathcal{I})$ and a random timestep $t$ (single integer) from $[0, 1000]$
    \State Obtain a noisy version of $x_0$, ($x_t$) using $\epsilon$ and $t$ to obtain $x_t$. Higher values of $t$ would indicate a noiser $x_t$.
    \State Concatenate $[x_0, c]$ in the channel dimension, repeating for each time index.
    \State Obtain $\nu$, a reparameterization of $\epsilon$
    \State Use the denoising model $\theta([x_0, c])$ to obtain $\nu_\theta$, a prediction of $\nu$
    \State Apply mean squared error to $\nu$ and $\nu_\theta$
    \State Take a gradient descent step
    \State Repeat from step 2
\end{algorithmic}
\end{algorithm}

\subsection{Model Tuning}
For training, we use the Adam optimizer, with hyperparameters described in Table~\ref{tab:hparams}. Due to time and computational constraints, we manually explored the hyperparameter space in a limited fashion. Resources permitting, a guided hyperparameter search would likely yield even better results.

% \usepackage{booktabs}
\begin{table}
\caption{Training Hyperparameters} \label{tab:hparams}
\centering
\begin{tabular}{@{}ll@{}}
\toprule
\textbf{Adam HyperParameters}       & Value  \\ \midrule
\multicolumn{1}{l}{Learning Rate}  & 0.0001       \\
\multicolumn{1}{l}{$\beta_1$}      & 0.9    \\
\multicolumn{1}{l}{$\beta_2$}      & 0.99   \\
\multicolumn{1}{l}{$\epsilon$}     & 1e-8   \\ \midrule
\textbf{Diffusion Hyperparameters}  &        \\ \midrule
\multicolumn{1}{l}{Sampling Steps} & 25    \\
\multicolumn{1}{l}{Noise Schedule} & Linear \\
\multicolumn{1}{l}{Loss type}      & L2     \\ \bottomrule
\end{tabular}
\end{table}

\section{Results and Analyses}
\subsection{Evaluating DiffESM}
In this section, we present a comprehensive evaluation of DiffESM's ability to generate month-long sequences of daily temperature or precipitation whose statistical characteristics closely match those of the target ESM output used for conditioning. For the large set of analyses detailed here, we use data from the IPSL ESM, under the RCP8.5 scenario for the years 2080-2100. Subsequent sections will address the performance over various 20-year time windows throughout the 21st century, as well as the emulator performance for a new scenario and a different ESM (CESM). 

After training DiffESM, we generate an ensemble of new daily sequences by conditioning on monthly averages from a held out ensemble member not seen during training. Note that this is not the intended deployment scenario, where monthly averages would instead come from a low temporal resolution emulator, but is part of the experimental design to facilitate evaluation of our samples.
We compare the generated value to the daily values of both the conditioning sequence (referred to as Held Out 1 or HO1) and a second held-out ensemble member (referred to as Held Out 2 or HO2). This approach allows us to assess how well the generated data matches the spatiotemporal characteristics of the conditioning data and how the variability in the generated ensemble compares to the inter-member variability of the target ESM. By comparing HO1 and HO2, we establish an oracle baseline to gauge the performance of our emulator. The discrepancies between these ensemble members provide a lower bound on the expected discrepancies between the generated data and HO2. Ideally, a larger number of ensemble members would be used to build a distribution of such differences, but the computational constraints that motivated the development of DiffESM also limit the number of independent realizations available from an ESM.

\begin{figure}
    \centering
    \includegraphics[width=\textwidth]{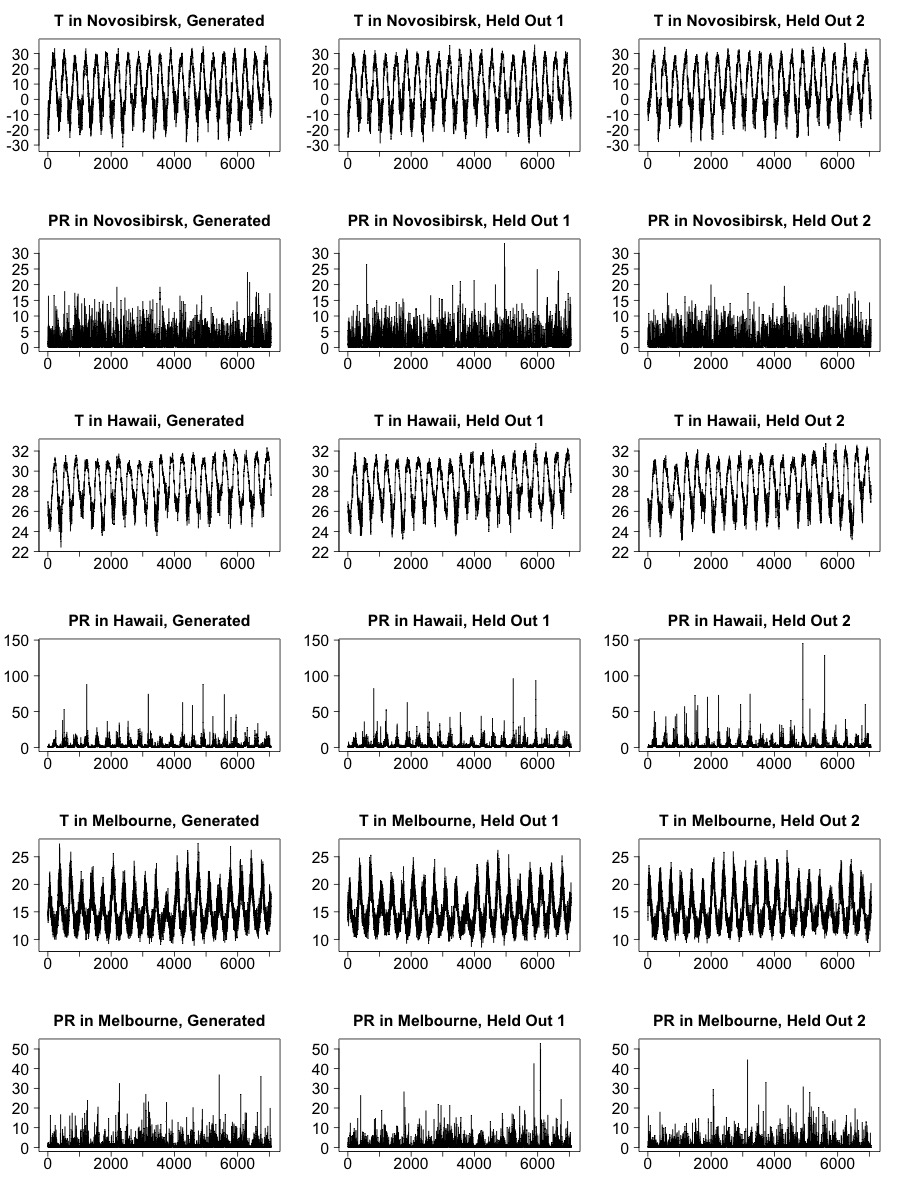}
    \caption{Time series of generated, Held Out 1 and Held Out 2 sets of daily temperature (T) and precipitation (PR) at the three locations, covering the period 2080-2100. The x-axis shows integer values indexing the days that span the period 2080-2100 (4 weeks per month). }
    \label{fig:time-series}
\end{figure}

\begin{figure}
    \centering
    \includegraphics[width=\textwidth]{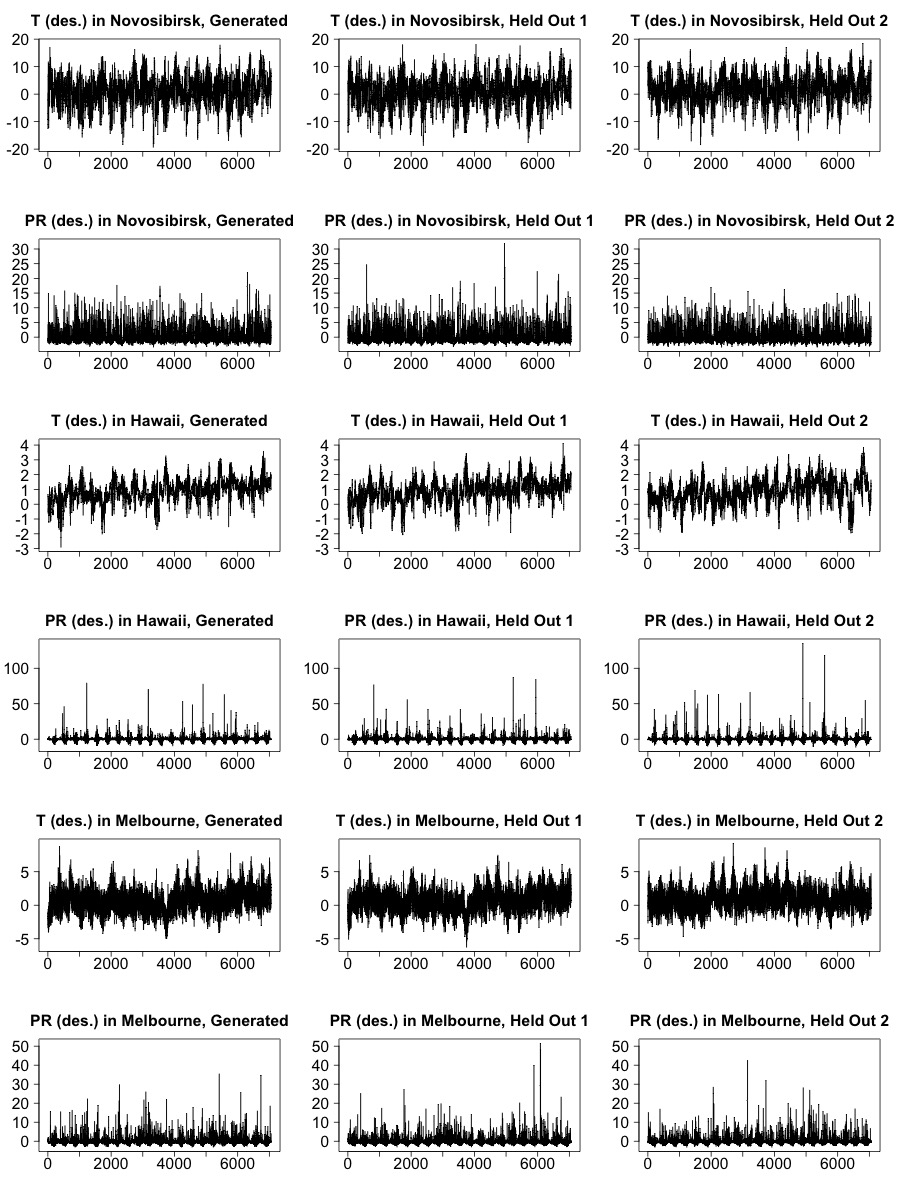}
    \caption{As Figure~\ref{fig:time-series} but after subtracting the seasonal cycle.}
    \label{fig:time-series-deseas}
\end{figure}

Figure~\ref{fig:time-series} presents time series of daily temperature and precipitation values for the generated data and the two held-out ensemble members at three example locations chosen to represent diverse climates. We extract the output of the ESM and the diffusion model at the three grid points closest to the big island of Hawaii, Melbourne (Australia) and Novosibirsk (Russia). The different ranges of temperature and precipitation values and the distinct characteristics of their seasonal cycles across the three locations are noteworthy. Visual inspection suggests that the generated data and the ESM data from HO1 and HO2 have indistinguishable behavior, but we analyze and document this in greater detail in the coming sections.
Figure~\ref{fig:time-series-deseas} displays the same time series after subtracting the seasonal cycle, confirming that the consistent behavior is not solely a result of emulating the mean monthly signal. 

\subsubsection{Spatial and Temporal Distributions}
\begin{figure}
    \centering
    \includegraphics[width=\textwidth]{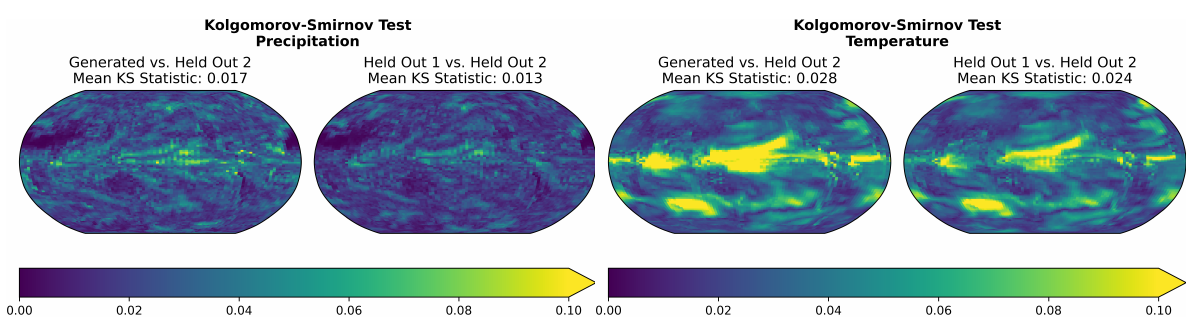}
    \caption{Global Kolgomorov-Smirnov Tests for Precipitation and Temperature. Values exceeding 0.10 are displayed as the same color.}
    \label{fig:default-ks}
\end{figure}

Figure~\ref{fig:default-ks} presents the result of the two-sample Kolgomorov-Smirnov (KS) test comparing the cumulative distribution functions (CDFs) of daily output at each grid-point (or pixel location in diffusion model terminology) for the generated data and the two held-out sets. To create this figure, we start with the three 20-year long daily time series of generated data and our two held-out members. For each pixel location, we perform a KS test comparing two empirical CDFs of daily temperature (or precipitation) estimated using the entire 20-year time series (approximately 7000 days), where the two CDFs are either those of Generated and HO2 or those of HO1 and HO2. The value of the test statistic in each location represents the maximum distance between the two empirical CDFs. The KS values in the plot comparing HO1 and HO2 gives us a measure of the natural variability of the model output's distribution across realizations with different initial conditions. We also perform the same KS calculation between the generated data and HO2.
Finally, to distill this information down to a single scalar value per comparison, we compute a mean ``KS value'' by averaging all the pixel values over the two maps (by a cosine-weighted average). These analyses are performed separately for temperature and precipitation data.

For precipitation, (Figure~\ref{fig:default-ks}), the two maps share many similar features globally, indicating that the generated data comes from a distribution across the globe very similar to that of the HO1 dataset (used to derive the monthly conditioning for the generated). Notably, the average KS score is only 0.004  higher on our generated vs. HO2 map, suggesting that the CDF of the generated precipitation data is only slightly more discrepant from the HO2 CDF than the CDF from HO1 is. Similar behavior is observed for temperature data. The comparability of the grid-point-level KS values between the two plots also demonstrates that the spatial characteristics of the generated data match those of the real data and are not those of a smoother field (which could be the output of a less accurate emulator trained on monthly means).

\begin{figure}
    \centering
    \includegraphics[width=\linewidth]{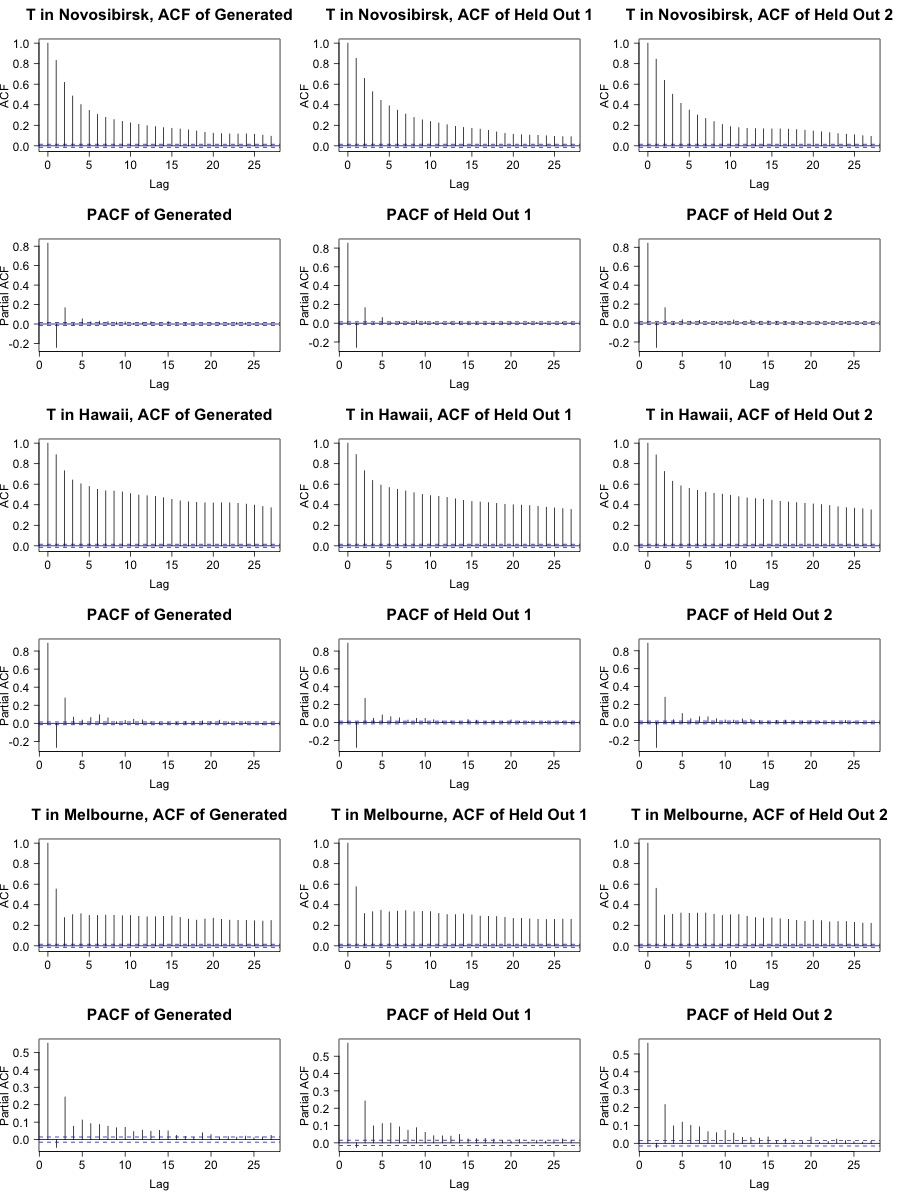}
    \caption{Autocorrelation and partial autocorrelation functions of daily time series of temperature (T) at the three locations (after subtracting the seasonal cycle). Generated data ACFs and PCFs along the left column can be compared to those of the Held Out 1 and Held Out 2 sets, along the middle and right column respectively. Each pair of rows corresponds to one of the three locations, with Novosibirsk at the top, Hawaii in the middle, and Melbourne at the bottom. }
    \label{fig:ACF_T}
\end{figure}

\begin{figure}
    \centering
    \includegraphics[width=\linewidth]{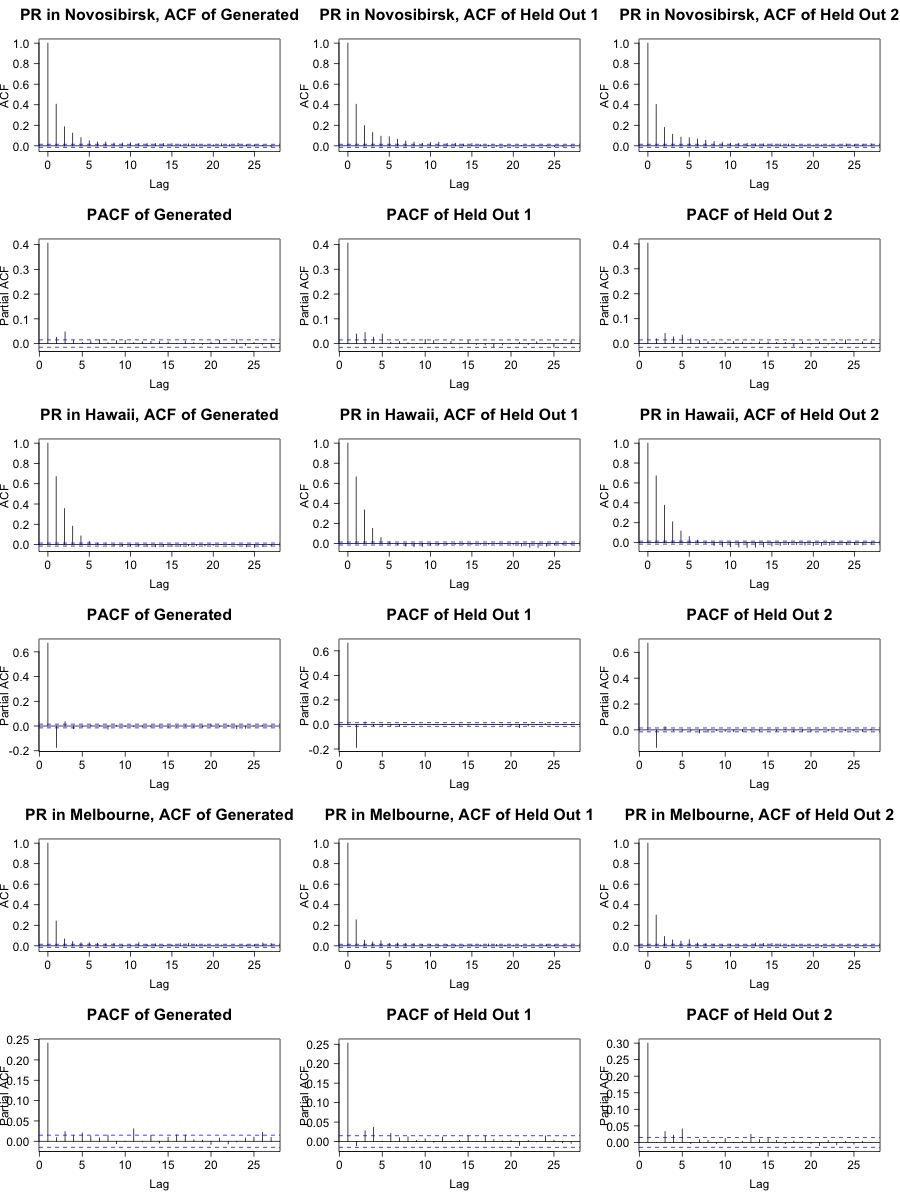}
    \caption{Like Fig. \ref{fig:ACF_T}, for daily precipitation time series.}
    \label{fig:ACF_Pr}
\end{figure}

To assess the temporal behavior of emulated versus ESM data, we examine the memory characteristics of each time series set. Figure~\ref{fig:ACF_T} and \ref{fig:ACF_Pr} contain autocorrelation (ACF) and partial autocorrelation (PACF) function plots for temperature and precipitation at the three locations whose time series are shown in Figure~\ref{fig:time-series} and~\ref{fig:time-series-deseas}. These plots represent the correlation between data points of the time series separated by an increasing number of lags (i.e., days, in our case up to 27 given the length of the emulated sequences). By eliminating the seasonal cycle each month is now treated as an independent and identically distributed set of daily quantities so we can estimate a single ACF (or PCF) using the statistical power that all our samples provide, rather than estimating an ACF/PCF pair for each month separately.
 The ACF behavior is affected by the lag-1 correlation dying off slowly and affecting subsequent lags. The PACF computes only the residual correlation at lag $n$ after accounting for correlations at lag 1 through $n-1$. ACF and PACF behavior is evaluated by comparing the overall shape of the former, and the number of significant ``spikes'' in the latter. Together these characteristics indicate the order of the Auto-regressive/Moving Average (ARMA) process \citep{boxjenkins} that generated the time series. 

The plots in Figure~\ref{fig:ACF_T} and~\ref{fig:ACF_Pr} confirm the consistency of the temporal structure of the generated, HO2 and HO1 datasets. It appears that the day-to-day memory of temperature and precipitation do not differ significantly between emulated and ESM data for all three locations, despite the range of behaviors shown by the (partial) autocorrelation functions. 


\subsubsection{Climate Metrics}\label{sssec:climate_metrics}
\begin{figure}
    \centering
    \includegraphics[width=\linewidth]{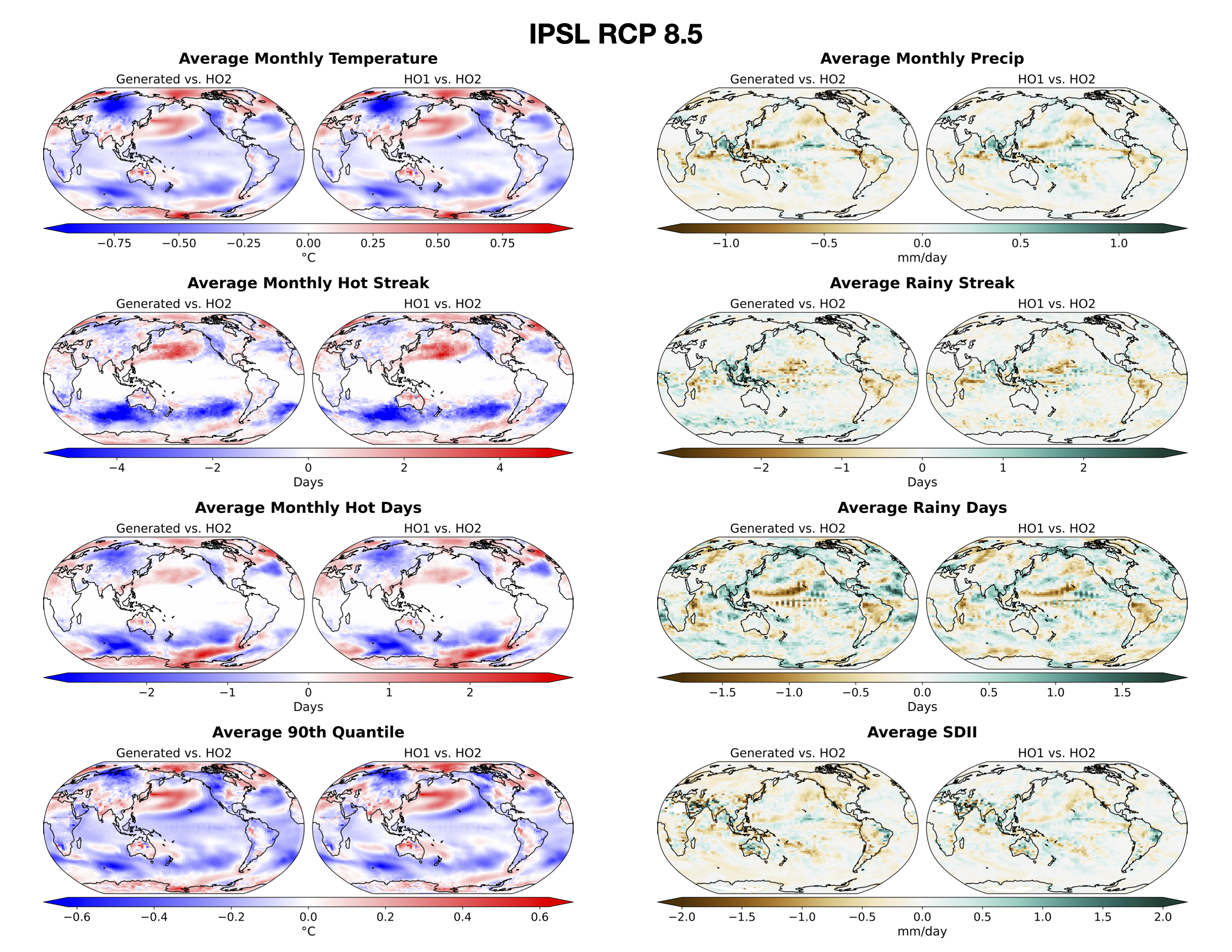}
    \caption{Relevant chosen metrics between generated set and HO1 conditioned on IPSL RCP8.5 runs from the years 2080-2100. See Section~\ref{sssec:climate_metrics} for  details on how these values are computed.}
    \label{fig:metric-figs}
\end{figure}

Daily data from ESMs is often used to derive metrics representative of extreme behavior, such as hot streaks, rainy or dry streaks, and the intensity of hot and rainy days. We choose a set of such metrics, each summarizing the daily behavior over a month. Table~\ref{table:metric-table} describes each metric computation, among which the simple daily intensity index (SDII) has been borrowed from the standard set of ETCCDI (Expert Team on Climate Change Detection and Indices)
 metrics~\citep{ETCCDI}. Figure~\ref{fig:metric-figs} shows DiffESM's performance on a range of these metrics. For each dataset and grid-point, the metric of interest is computed for each month from 2080-2100. We then average over all months for HO2 to produce an HO2 map and over all months in HO1 to produce an HO1 map. Subtracting the HO2 map from the HO1 map provides a baseline for the level of internal variability between two realizations from the same ESM. We compute the same difference map between the generated and HO2 sets and compare it to this baseline.

% \usepackage{booktabs}
% \usepackage{graphicx}
\begin{table}
\caption{Description of climate metric calculations}
\label{table:metric-table}
\centering
\resizebox{\textwidth}{!}{%
\begin{tabular}{@{}ll@{}}
\toprule
\multicolumn{1}{c}{\textbf{Metric}}    & \multicolumn{1}{c}{\textbf{Description}}                                            \\ \midrule
\multicolumn{2}{c}{\textbf{Temperature}}                                                                                     \\ \midrule
\textbf{Average Monthly Temperature}   & The average of the daily temperature values within the month                        \\ \midrule
\textbf{Average Monthly Hot Streak} &
  \begin{tabular}[c]{@{}l@{}}The longest consecutive number of days with daily temperature values\\ above a precomputed 90th quantile threshold value (threshold computed from a reference\\ period in 1960-1990)\end{tabular} \\ \midrule
\textbf{Average Monthly Hot Days} &
  \begin{tabular}[c]{@{}l@{}}The total number of days within a month with daily temperature values above\\ the precomputed 90th quantile threshold value.\end{tabular} \\ \midrule
\textbf{Average 90th Quantile}         & The average temperature on days that exceed the precomputed 90th quantile.          \\ \midrule
\multicolumn{2}{c}{\textbf{Precipitation}}                                                                                   \\ \midrule
\textbf{Average Monthly Precipitation} & The average of the daily rainfall values within the month (mm/day)                  \\ \midrule
\textbf{Average SDII} &
  \begin{tabular}[c]{@{}l@{}}The sum of rainfall on days exceeding 1mm/day divided by the total number of days exceeding\\ 1 mm/day\end{tabular} \\ \midrule
\textbf{Average Rainy Streak}          & The longest consecutive number of days within a month exceeding 1mm/day of rainfall \\ \midrule
\textbf{Average Rainy Days}            & The total number of days within a month with rainfall exceeding 1mm/day.            \\ \bottomrule
\end{tabular}%
}

\end{table}

For both temperature and precipitation data, Figure~\ref{fig:metric-figs}, the difference plots for generated vs. HO2 and HO1 vs. HO2 are similar. DiffESM matches HO1's performance globally (when averaging values over all grid points) and captures many of the same spatial patterns in terms of the sign and magnitude of the differences over both land and oceans. Additionally, metrics that incorporate temporal structure such as rainy streaks or hot streaks demonstrate a high level of both spatial and temporal agreement between DiffESM and the conditioning dataset.

\begin{figure}
    \centering
    \includegraphics[width=1\linewidth]{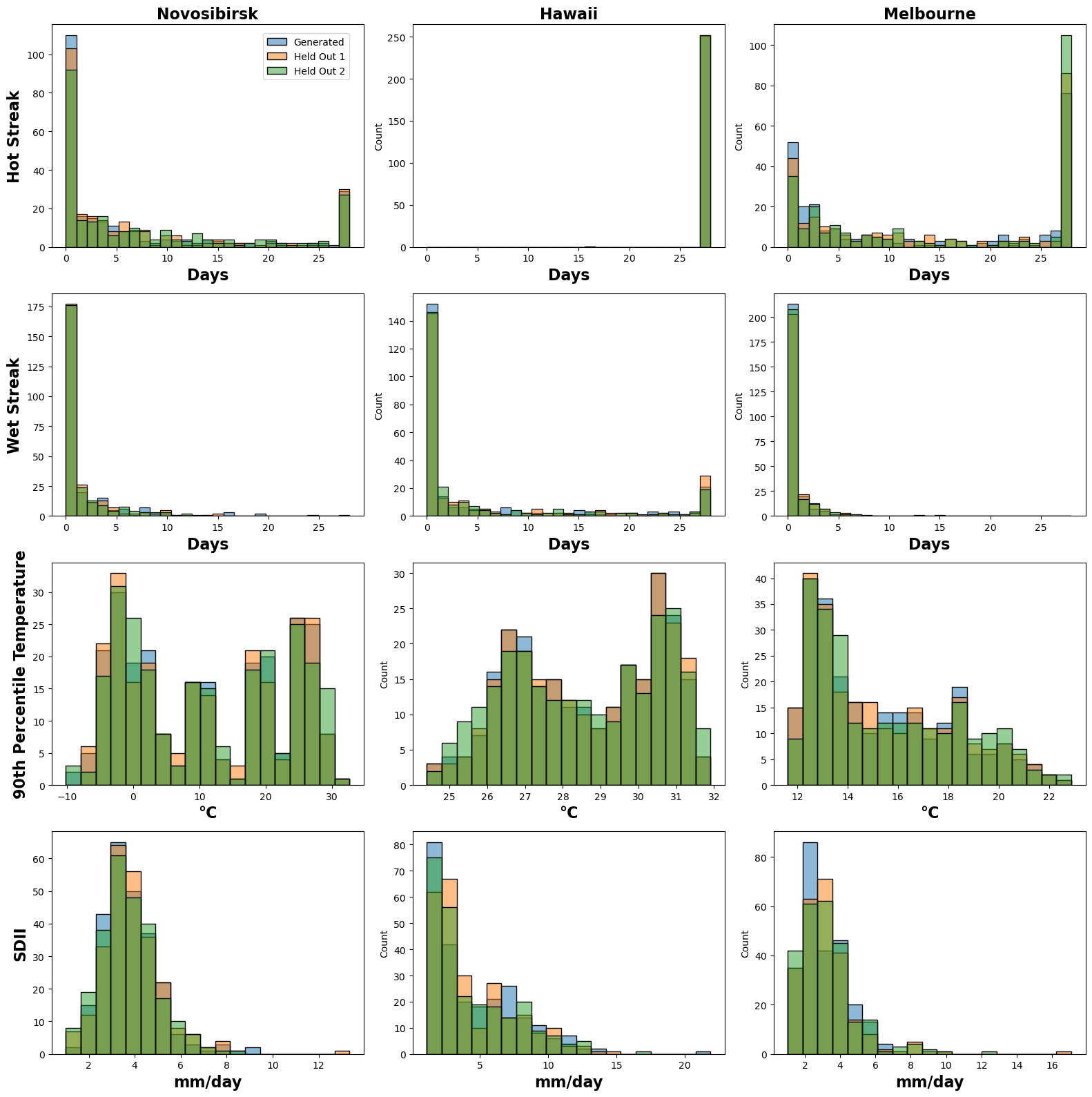}
    \caption{Histogram of metric performance at Melbourne, Hawaii, and Novosibirsk for the years 2080-2100. For metrics where ``days" are the unit, bins were chosen to be a width of 1 day. For all other metrics, twenty bins were chosen for visual clarity.}
    \label{fig:location_specific_metrics}
\end{figure}

For a more fine-grained evaluation of DiffESM's performance on each metric, Figure \ref{fig:location_specific_metrics} presents histograms of the results over time for the grid-points closest to Novosibirsk, Hawaii, and Melbourne. These histograms show that, when looking at metrics from specific locations over 2080-2100, our generated results are consistent with both HO1, used for conditioning, and HO2, the independent ensemble member.

\subsubsection{Variability} 
Some generative methods are known to suffer from ``mode collapse'' \citep{mode-collapse}, a tendency to generate only a small subset of the data distribution, resulting in samples with little internal variability. We have observed this issue in in the past in the context of ESM emulation with GANs. To ensure that DiffESM does not exhibit the same limitation, we assess the variability of the samples it produces.

\begin{figure}
    \centering
    \includegraphics[width=\linewidth]{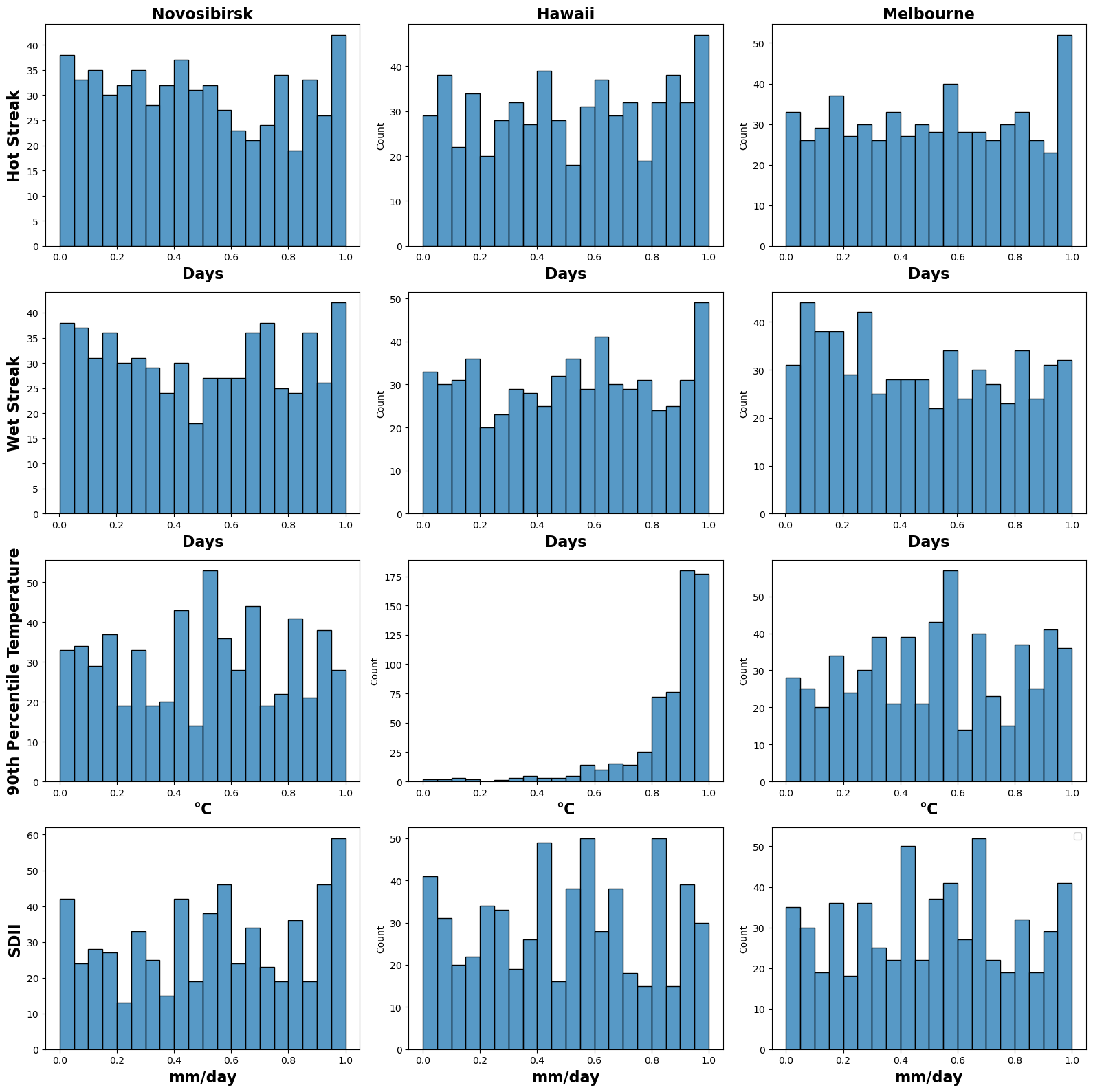}
    \caption{Location Specific Rank Histograms 2050-2100. Twenty bins were chosen for visual clarity and consistency of the results.}
    \label{fig:variability}
\end{figure}

Figure~\ref{fig:variability} presents the results of this assessment. For this analysis, we evaluate on the RCP8.5 scenario for our IPSL trained model, over the years 2050-2100. For each location, we compute the target metric on both the daily values of the HO1 dataset and 30 generated ensemble members. We then calculate the rank of the HO1 metric result within the 30-member generated ensemble for each location and month, yielding 612 fractional ranks (51 years $\times$ 12 months). These fractional ranks are plotted as histograms, which are designed to visually evaluate the``spread" of our generated ensemble. A ``U''-shaped histograms indicates an overly tight cluster, a dome-shaped histogram suggests too wide a  spread, and a skewed histogram indicates bias, either under- or over-predicting. A flat histogram is the indication of a sample whose distribution is the same as that of the quantity it is designed to emulate. 

The only performance that shows room for improvement is related to DiffESM consistently under-predicting the temperature of hot days in Hawaii, as the histogram with the peak to its right edge suggests (the truth has the highest rank in a majority of cases). Further investigation reveals that the average bias between the generated ensemble mean and the HO1 metric (for 90th percentile temperature in Hawaii) is -0.044, while the average mean squared error is 0.002. Despite the generated metric being extremely close to the HO1 metric, the consistent bias and small standard deviation of 0.031 for the generated ensemble lead to a rank histogram that displays consistent under-prediction. However, for other metric-location pairs, the histograms display relatively uniform behavior, indicating that the HO1 result falls evenly within the generated ensemble. This demonstrates that in the great majority of cases, the generated ensemble produces sufficient variability to capture the realm of possibilities for the true value, and does not ``overdo it,'' so that the truth has similar chances to fall anywhere within the sample. The uniform histograms also suggest that DiffESM does not suffer from the mode collapse issue observed in GANs, as it generates samples with adequate internal variability. A sample suffering from mode collapse would show a U shape or a skewed shape. 
 
\subsection{Analysis: Performance Across RCPs}
In this section we present analyses of DiffESM's performance across a wide range of forcing scenarios and years. For brevity, we only include a subset of the analyses described above that characterize the performance of our model both spatially and temporally.

\begin{sidewaysfigure}
    \centering
     \includegraphics[width=\linewidth]{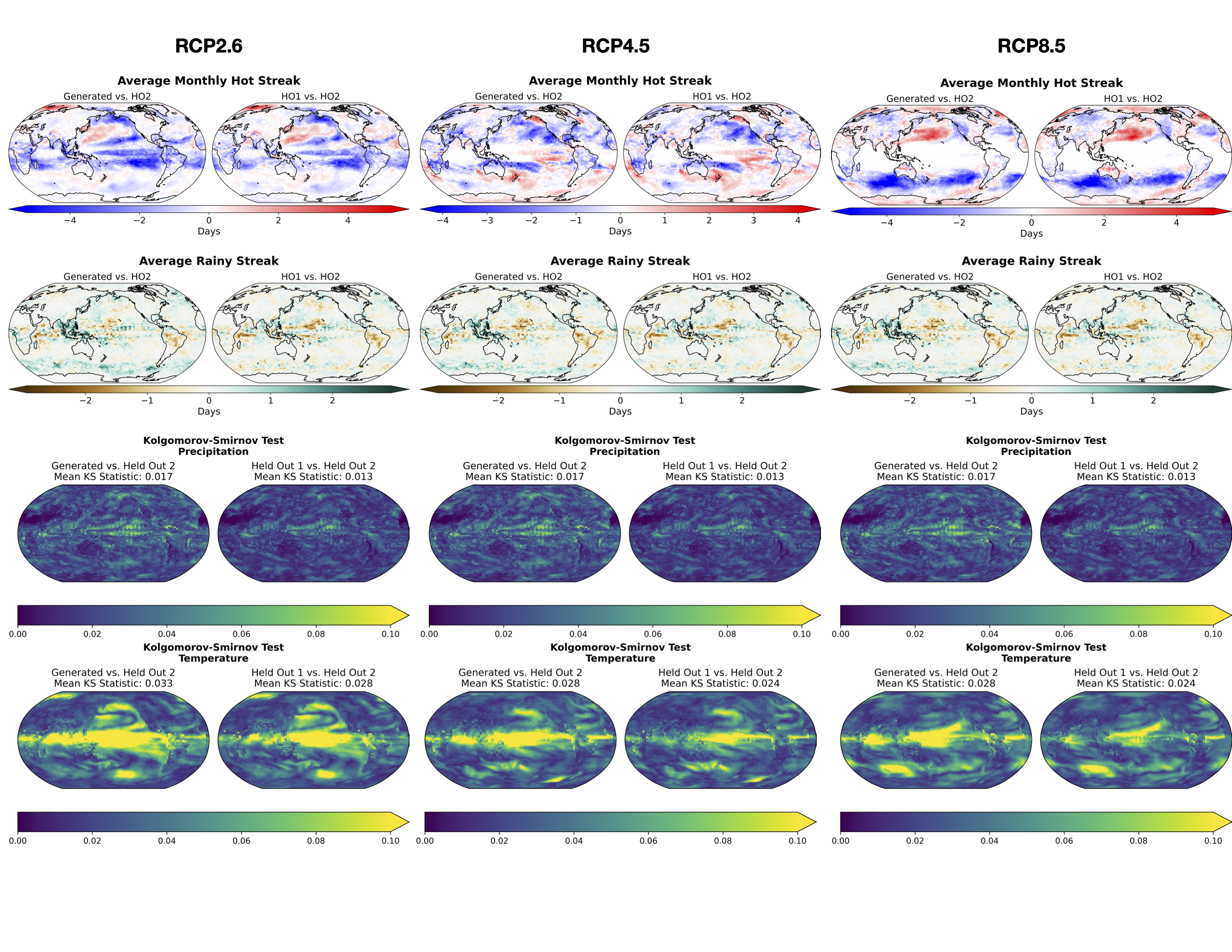}
    \caption{IPSL Model, evaluated under multiple forcing scenarios}
    \label{fig:rcp-varied}
\end{sidewaysfigure}

Figure~\ref{fig:rcp-varied} displays our IPSL-trained emulator generating data for different RCP forcing scenarios. DiffESM, trained on data from RCP8.5 (the highest emissions scenario available, covering the largest range of radiative forcings along the 21$^{\mbox{st}}$ century), is assessed on its ability to emulate the distribution of previously unseen, lower forcing scenarios, RCP2.6 and RCP4.5. The experimental setup is the same as before: for each of the new RCPs, we target the emulation of the years 2080-2100; we generate the 20 years of data with our emulator conditioned on a held-out realization, HO1, and compare its performance against a second realization, HO2 (we note that the labels are here used for consistency, but the concept of held-out realization in the case of RCPs other than RCP8.5 is redundant, as the diffusion model is trained only on RCP8.5). Our analyses show that DiffESM exhibits performances under the two lower scenarios comparable to the one assessed within RCP8.5, in each of the chosen climate metrics. For example, the KS test demonstrates that daily sequences emulated by the diffusion model closely match the underlying distributions of daily temperature and precipitation from the ESM run under these scenarios, despite having never encountered these emissions scenarios during training. We attribute this to the fact that the ESM output (daily temperature and precipitation) does not exhibit path-dependent (or long-memory) behavior, i.e., the shape of the scenario leading to a given month to emulate is not relevant, once that month is used for conditioning (in statistical speech,  the month would be defined a sufficient statistic). Thus, conditioning on a map of average temperature or precipitation is sufficient to recreate the correct behavior, provided that the emulator has been trained on output reflecting those types of mean maps, independently of the scenario along which they were reached.

\subsection{Analysis: Performance Across Time Periods}
\begin{sidewaysfigure}
    \centering
    \includegraphics[width=\linewidth]{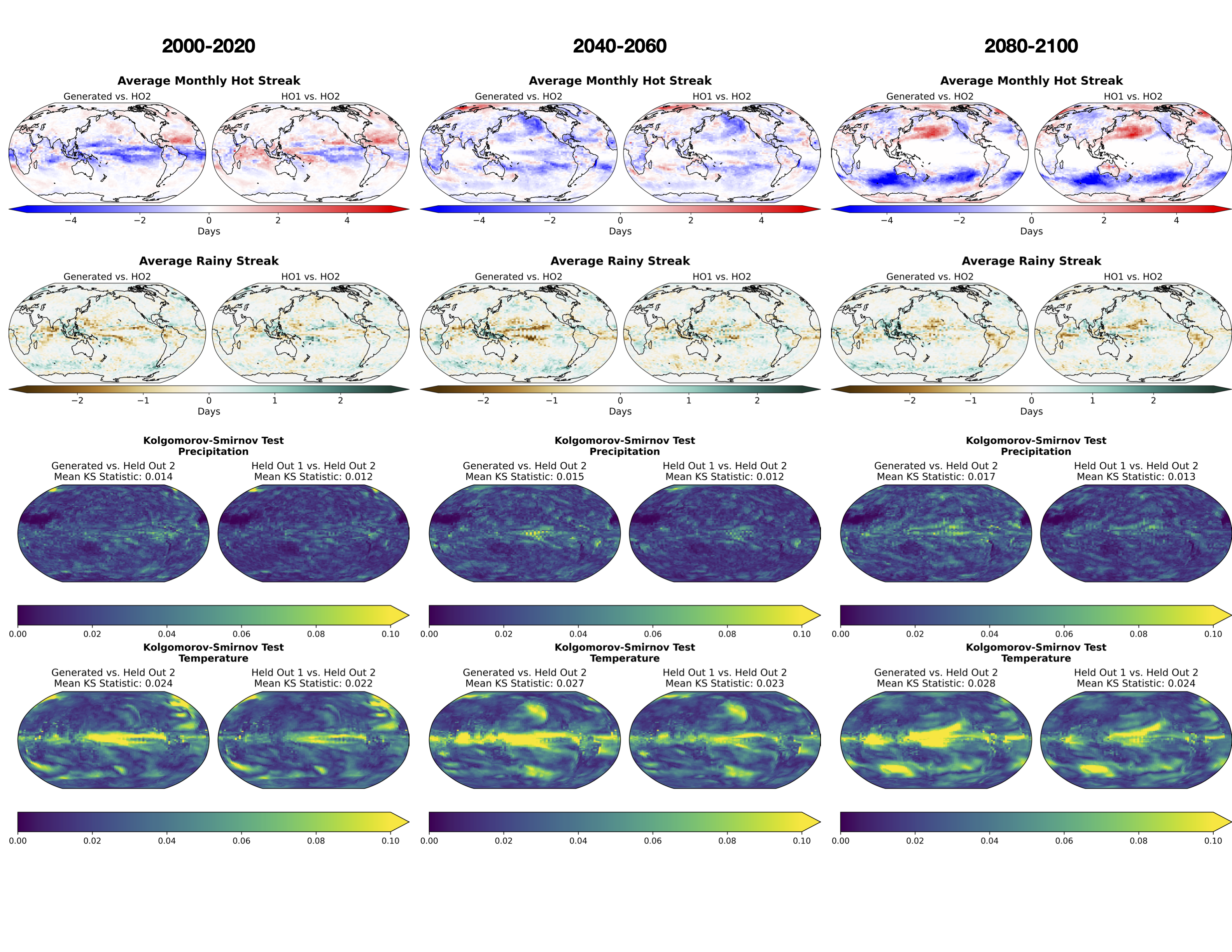}
    \caption{Performance of DiffESM across multiple timespans}
    \label{fig:time-spans}
\end{sidewaysfigure}

Here, we analyze DiffESM's performance across different time periods, specifically three 20-year windows: 2000-2020, 2040-2060, 2080-2100 (early century, mid-century and late-century). Figure~\ref{fig:time-spans} shows that the results remain mostly consistent across multiple time periods, differentiating the outcome of the transient scenario across the time windows.

\subsection{Analysis: Generalization of Emulator Approach to a New ESM}
\begin{figure}
    \centering
    \includegraphics[width=\linewidth]{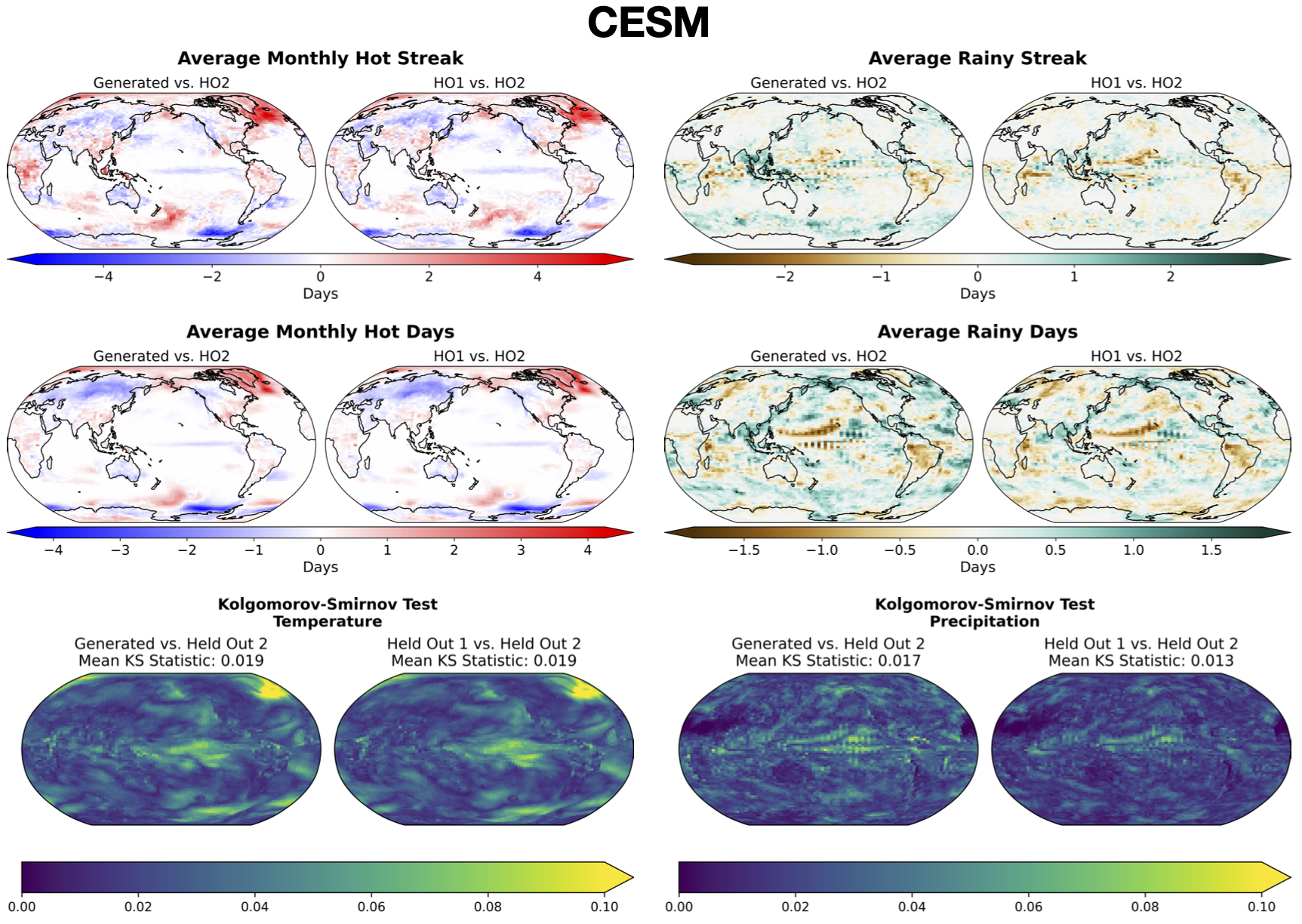}
    \caption{Performance of a model trained on CESM data, evaluated on the CESM dataset}
    \label{fig:ESM varied}
\end{figure}

Although our primary focus has been on the IPSL ESM, we demonstrate that the emulation process can be replicated on another ESM; namely CESM. Figure~\ref{fig:ESM varied} demonstrates DiffESM's on the CESM dataset. Specifically, we train a new DiffESM model on CESM data and analyze its performance, finding that it closely matches the spatial and temporal distribution of the ESM. According to both the KS maps, and the maps of metric-differences, DiffESM performs similarly when trained and evaluated on CESM data, compared to IPSL data. This demonstrates that the approach of training a diffusion model to downscale monthly averages to daily values can be effectively applied for multiple ESMs.

\subsection{Analysis: Effect of Training Set Size on Performance}
In this section, we investigate the impact of the number of realizations used for training DiffESM on its performance. We train three separate DiffESM models using one, two, or three realizations from the IPSL ESM under the RCP8.5 scenario for the years 2080-2100. The models are then evaluated using the same set of analyses described in previous sections.

\begin{sidewaysfigure}
    \centering
    \includegraphics[width=\linewidth]{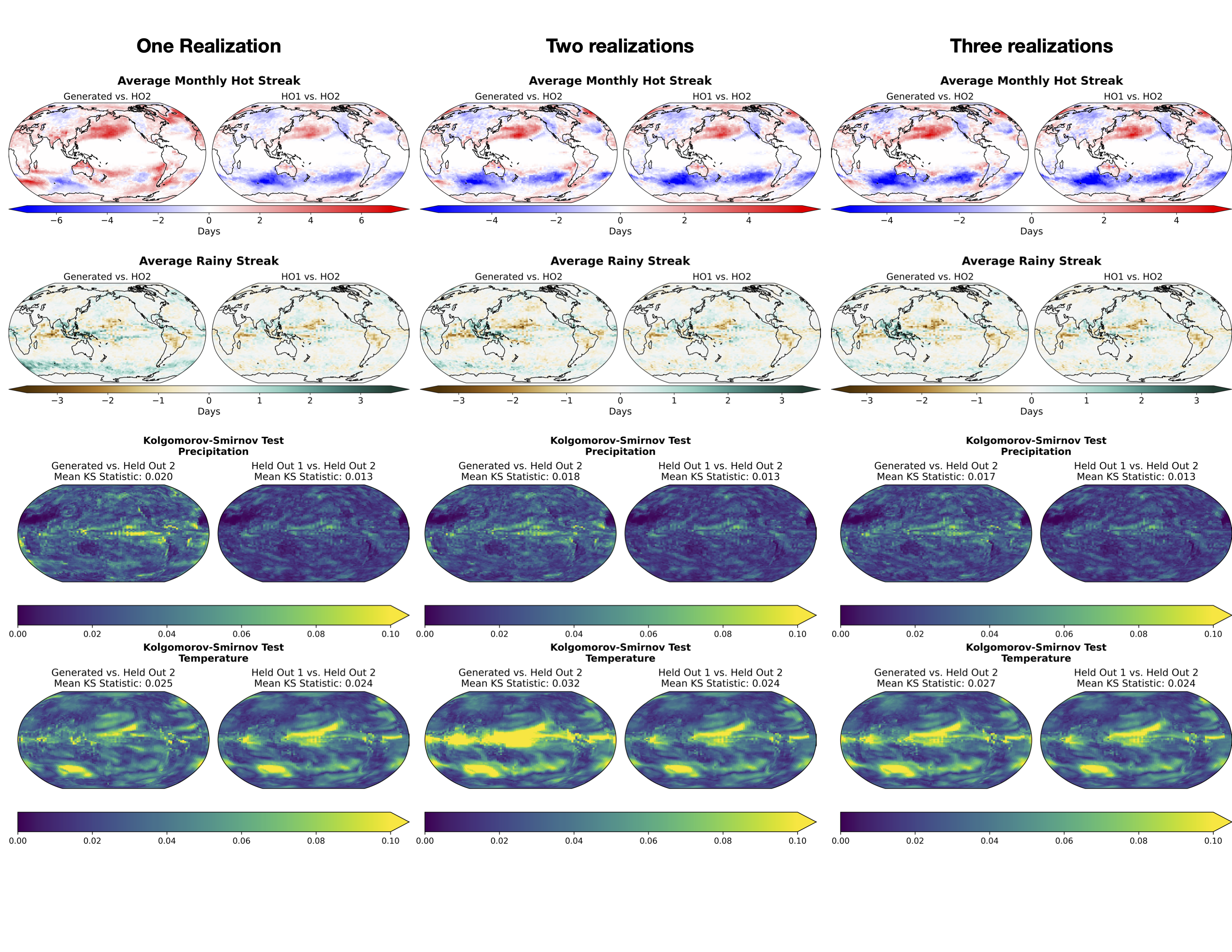}
    \caption{Performance of DiffESM for different training set sizes.}
    \label{fig:training-size}
\end{sidewaysfigure}

Figure~\ref{fig:training-size} presents the results of this comparison. The difference plots show that, although there are subtle changes between each model, the generated vs. HO2 map is still highly similar to the HO1 vs. HO2 map for each metric displayed. Additionally, the KS tests show that there is a slight degradation in performance for precipitation with less training data, but overall DiffESM is able to capture the spatial characteristics of both variables, regardless of training set size. 

These results suggest that DiffESM's performance is relatively robust to the size of the training set, at least within the range of one to three realizations. This finding is particularly important, given the computational constraints that limit the number of realizations available from ESMs and the different resources available across modeling centers.

\section{Conclusions and Future Work}
In this paper, we have demonstrated the capability of DiffESM, a conditional video diffusion model, to emulate ESM output of daily temperature and precipitation conditioned on monthly means, also for climate scenarios unseen during training.
We observe that the samples produced by DiffESM are comparable to those of ESMs in some fundamental characteristics, such as temporal correlation and spatial behavior, and in several extreme-relevant metrics, such as frequency and spatial distribution of hot streaks or dry spells, and intensity of precipitation during extremely wet days. In fact, we have shown that for many performance metrics, the emulator errors (the differences from the ESM output it targeted to emulate) are similar to differences between different realization from the ESM itself,  i.e., comparable to internal variability.  The ability to generate such simulations in concert with a monthly mean emulator (a more commonly available type, at this time)  could significantly enhance our ability to characterize the risks from extreme weather events under various future climate scenarios. Of course the limitations of global climate models in representing the longer tails and more extreme extremes of climate variables remain valid for our emulator. The climate extreme metrics we have evaluated are similar in definition to some ETCDDI indices, which have been constructed with those limitations in mind, and extensively validated over several generation of models by now, ~\citep{sillmann, Seneviratneetal}. Another -- more pragmatic -- use of emulation of daily quantities from monthly means that we foresee, as we work towards improving the emulator performance could be to decrease the cost of archiving and handling ESM daily output, which is becoming increasingly high due to ESMs' higher and higher resolution.

There are numerous directions for future work, beyond what we here propose as our first exploration of the potential of diffusion models for climate model output emulation. One promising area would be to integrate multiple variables into a single diffusion model, since modeling the correlation between, for example, temperature and precipitation would allow for investigation of co-occurrent phenomena, such as the interaction between temperature and precipitation in creating extreme hot and dry conditions.  We plan to bypass the limit of the 28-day length unit by autoregressively generating months with conditioning to promote continuity at month boundaries, and we are exploring conditioning on longer averages (e.g., annual means cascading temporal downscaling). Despite the speed advantages over ESMs, the diffusion models could themselves be further sped up using sampling techniques such as progressive distillation \citep{variational}. Lastly, while the work reported in this manuscript emulates two ESMs and evaluates the emulated output on three scenarios (two unseen in training), we plan to replicate these findings over many more ESMs and scenarios to further evaluate the promise of these techniques.

\section{Open Research}

The code used for training and evaluating the models in the study are available via an MIT license at \citep{Seth_Bassetti_and_Brian_Hutchinson_and_Claudia_Tebaldi_and_Ben_Kravitz_DiffESM_2023}.
The data used to train our models was obtained from the Earth System Grid, part of the CMIP5 archive, which can be obtained from the Earth System Grid Federation database: \url{https://aims2.llnl.gov/search/cmip5/}.

%%%%%%%%%%%%%%%%%%%%%%%%%%%%%%%%%%%%%%%%%%%%%%%

\acknowledgments
This research was supported by the U.S. Department of Energy, Office of Science, as part of research in MultiSector Dynamics, Earth and Environmental System Modeling Program. The Pacific Northwest National Laboratory is operated for DOE by Battelle Memorial Institute under contract DE-AC05-76RL01830. Support for BK was provided in part by the National Science Foundation through agreement SES-1754740 and the Indiana University Environmental Resilience Institute. The views and opinions expressed in this paper are those of the authors alone. 

Authors would also like to acknowledge the World Climate Research Programme's Working Group on Coupled Modelling, which is responsible for CMIP, and we thank the climate modeling groups
at the National Center for Atmospheric Research (USA) and Institut Pierre Simon Laplace
(France) for producing and making available their model output. For CMIP the U.S. Department of Energy's Program for Climate Model Diagnosis and Intercomparison provides coordinating support and led development of software infrastructure in partnership with the Global Organization for Earth System Science Portals.

Finally, the authors would like to thank the Nvidia Corporation for the donation of GPUs used in this research.

%% ------------------------------------------------------------------------ %%
%% References and Citations

%%%%%%%%%%%%%%%%%%%%%%%%%%%%%%%%%%%%%%%%%%%%%%%
%
\bibliography{agusample}
%
% don't specify bibliographystyle

% In the References section, citep the data/software described in the Availability Statement (this includes primary and processed data used for your research). For details on data/software citation as well as examples, see the Data & Software Citation section of the Data & Software for Authors guidance
% https://www.agu.org/Publish-with-AGU/Publish/Author-Resources/Data-and-Software-for-Authors#citation

%%%%%%%%%%%%%%%%%%%%%%%%%%%%%%%%%%%%%%%%%%%%%%%

%\bibliography{enter your bibtex bibliography filename here}

%Reference citation instructions and examples:
%
% Please use ONLY \citep and \citepA for reference citations.
% \citep for parenthetical references
% ...as shown in recent studies (Simpson et al., 2019)
% \citepA for in-text citations
% ...Simpson et al. (2019) have shown...
%
%
%...as shown by \citepA{jskilby}.
%...as shown by \citepA{lewin76}, \citeA{carson86}, \citepA{bartoldy02}, and \citeA{rinaldi03}.
%...has been shown \cite{jskilbye}.
%...has been shown \cite{lewin76,carson86,bartoldy02,rinaldi03}.
%... \cite <i.e.>[]{lewin76,carson86,bartoldy02,rinaldi03}.
%...has been shown by \cite <e.g.,>[and others]{lewin76}.
%
% apacite uses < > for prenotes and [ ] for postnotes
% DO NOT use other cite commands (e.g., \citet, \citep, \citeyear, \citealp, etc.).
% \nocite is okay to use to add references from your Supporting Information
%

\end{document}